# Detecting gamma-ray bursts with the Pierre Auger Observatory using the single particle technique

D. Allard, E. Parizot, X. Bertou, J. Beatty, M. Du Vernois, D. Nitz, G. Rodriguez
*for the Pierre Auger Collaboration*
Presenter: Denis Allard (denis@oddjob.uchicago.edu), arg-bertou-X-abs2-og24-poster

During the past ten years, gamma-ray Bursts (GRB) have been extensively studied in the keV-MeV energy range but the higher energy emission still remains mysterious. Ground based observatories have the possibility to investigate energy range around one GeV using the "single particle technique". The aim of the present study is to investigate the capability of the Pierre Auger Observatory to detect the high energy emission of GRBs with such a technique. According to the detector response to photon showers around one GeV, and making reasonable assumptions about the high energy emission of GRBs, we show that the Pierre Auger Observatory is a competitive instrument for this technique, and that water tanks are very promising detectors for the single particle technique.

## 1. Introduction

Instruments such as BATSE (of the Compton Gamma-Ray Observatory) and BEPPO-Sax have studied GRBs prompt emission in great detail between 20 keV and 2 MeV. The emission in this energy range is understood with the internal shocks model [1] mainly as the synchrotron emission of electrons accelerated by mildly relativistic shocks. Higher energy photons, during the prompt emission, have also been seen by EGRET (up to 2 GeV for GRB940217), but the high energy emission (around one GeV) has been poorly investigated, both experimentally and theoretically. Learning more about the high energy emission of GRBs is one of the missions of the future satellite GLAST but it is also possible for ground based observatories to contribute. In this case, the detection method used is called "the single particle technique" and it relies on the fact that around one GeV, photon showers give only isolated particles at ground level and are thus impossible to reconstruct. But, if the high energy emission of GRBs is sufficiently intense, the isolated particles at the ground could lead to a significant excess when compared with the background fluctuations. This method has already been investigated by S. Vernetto [2] and used for several experiments mainly using scintillators or RPC (for instance INCA in Bolivia [3] and ARGO YBJ in Tibet [4]).

The Pierre Auger Observatory [5], is a 3000 km$^2$ ground based detector located in Malargüe (Argentina) at 1400 m above the see level and dedicated to ultra energetic cosmic ray detection. The Pierre Auger Observatory surface detector (SD) is composed of 1600 Cherenkov water tanks of 12 m$^3$ and 3 photomultipliers are placed on the top of each tank. The detection of ground particles of air showers is based on the Cherenkov light they emit when passing through the water tank. These detectors are sensitive to the photon component of air showers [6](whereas scintillators are mainly sensitive to the $e^+e^-$ component). As photons are the most numerous particles at ground level and as the SD has a huge detection area for the single particle technique (1600 × 10 m$^2$), it is worth investigating the possibility to detect the high energy emission of GRBs with this observatory. For this purpose, we first determine the main properties of air showers around one GeV using the CORSIKA air shower simulation and study the detector response to ground particles to estimate the SD capability to detect the high energy emission of GRBs, taking into account observational and theoretical constraints.

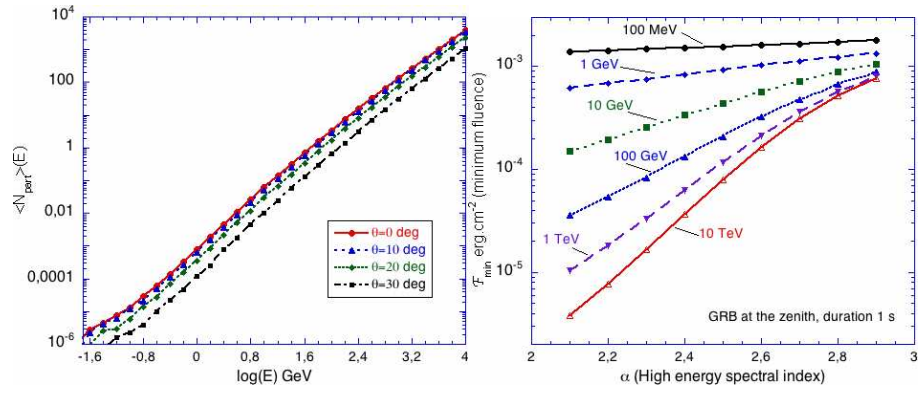

**Figure 1.** Left: Mean number of detected particles in a tank as a function of the photon primary energy for different zenith angles. Right: Minimum fluence between 10 MeV and $E_{max}$ for a GRB to be observable as a function of the spectral index.

## 2. GeV photon air showers

To determine the main characteristics of photon air showers around one GeV, we simulated 1.4 billion showers with CORSIKA [7] between 10 MeV and 10 TeV for zenith angles between $0°$ and $30°$. We then estimated the evolution of the number and the type of particles at ground level, their energy spectrum and angular distribution as a function of the primary photon energy and zenith angle.

Concerning the number of particles, the results of the simulation show that the number of particles increase with the energy as a power law (going from $2 \times 10^{-5}$ at 100 MeV to 6000 at 10 TeV for a zenith angle of $0°$). At a given primary energy, this number decreases with the zenith angle due to a much longer path in the atmosphere. The number of photons represents more than 90% of the total amount of particles at ground level, whatever the primary energy. Note that a muonic component appears above 1 GeV but the amount of muons at ground is almost negligible compared with electromagnetic particles. The spectrum of $e^+e^-$ and photons at ground level are very similar and most of the particles are in 1-30 MeV range with a high energy tail above 100 MeV (a very small part of the total number) for high energy showers. We note that these spectra are nearly independent of the primary photon energy (except on the fact that the high energy tail is cut for low energy primaries). The angular distribution of particles shows that secondary photons and $e^+e^-$ are very well correlated with the primary direction. Despite this low angular divergence, multiple hits of particles produced by the same shower in a given tank are only relevant above 1 TeV.

## 3. Response of the detectors to photon showers

Knowing the characteristics of particles at ground level, we estimated the detector response to these particles. We simulated photons and electrons between 1 MeV and 1 GeV with incident directions between $0°$ and $60°$, using GEANT4 [8] adapted to the need of the SD to determine the probability to trigger (i.e to be counted) as a function of the incoming particle energy and incidence direction for different trigger strategies (the trigger strategy we finally choose is the one of the SD scalers, i.e, a 4 ADC counts threshold required for at least 1 PMT). The resulting probabilities can be convoluted with the particle content at ground level of photon showers to determine the mean number of particles detected at ground as a function of the primary photon energy and of the zenith angle. The result of the calculation is shown in figure 1 (left). Most of the detected particles are photons (around 90%) and the mean number of detected particles as a function of energy increases as a power law (the index is almost the same as for the number of particles at ground level). It is important to note that the use of the single particle technique makes the detection completely independent of the size of the tanks and of their granularity. The only relevant quantity is the detectable particles density at ground which is the convolution of the gamma-ray flux at the top of the atmosphere, the development of showers in the atmosphere and the detector response.

## 4. What can be seen with the Pierre Auger Observatory?

The condition for a GRB to be detected is that its counting rate at ground level summed over all the tanks has to be significantly higher than the background fluctuations. Let us consider a GRB emitting a luminosity $L$ (erg s$^{-1}$) in an energy range $[E_{\min}, E_{\max}]$ at a redshift $z$. The flux at the Earth and the implied detection condition are then given by :

$$\Phi = \frac{L}{4\pi d_L^2(z)} = \int_{\frac{E_{min}}{1+z}}^{\frac{E_{max}}{1+z}} E.n(E)\, dE \Longrightarrow \int_{\frac{E_{min}}{1+z}}^{\frac{E_{max}}{1+z}} P_{\text{detec}}(E).n(E)\, dE \geq \frac{s}{S_\theta A}\sqrt{\frac{B}{N\Delta t}} \quad (1)$$

where $d_L$ is the luminosity distance of the source, $n(E)$ the differential spectrum of photons (s$^{-1}$ cm$^{-2}$ erg$^{-1}$), $P_{\text{detec}}(E)$ the mean number of triggering particles from a shower induced by a photon of energy $E$, $s$ the desired statistical significance (in the following we will take $s$=5), $A$ the area of a tank, $N$ the number of tanks, $B$ the background counting rate (for the trigger strategy we choose, the background is around 3.8 kHz/tank), $\Delta t$ the duration of the particle flux, and $S_\theta$ is a geometric factor depending on the zenith angle of the gamma. An interesting point to note is that the detection threshold (s$^{-1}$ m$^{-2}$) evolves as $(\Delta t N)^{-1/2}$ which means that the more tanks we have and the longer the particle flux lasts, the weaker the flux can be.

As a first estimate, we can make a very simple and naive assumption that the high energy emission of GRBs is the simple extrapolation of the spectra seen by BATSE. These spectra scale as a power law above the peak energy with an index typically between 2 and 3 (the mean value is around 2.2, and no high energy cut-off has been observed [9, 10]). We have then estimated, making the assumption that high energy spectra scale as $n(E) = \kappa E^{-\beta}$ (with $2 \leq \beta \leq 3$ and $\kappa$ spectrum normalization constant), the minimal fluence (erg cm$^{-2}$) between 10 MeV and $E_{\max}$ that could be detected by the SD as a function of $\beta$ and for different values of $E_{\max}$. This estimate can be done straightforwardly using equation 1 to calculate the normalization constant and then deduce the fluence. Figure 1 (right panel), shows the results of this calculation. However, the BATSE catalogs [9] are composed of observed fluences (in the 20 keV - 2 MeV range) typically between $10^{-6}$ and $5 \times 10^{-4}$ erg cm$^{-2}$. One can see on the figure 1 (right) that we obtain realistic fluences above 10 MeV in the case of an energy cutoff above 1-10 GeV (for a duration $\Delta t$, fluences have to be multiplied by a factor $\sqrt{\Delta t}$). Thus we can estimate that the high energy emission of bright and/or nearby bursts with a low zenith angle can be at least constrained and if not observed in the case of a high energy cutoff well above 1 GeV.

The event rate is difficult to infer but we can compare these minimum fluences for $E_{\max} = 100$ GeV with predictions made in [2] for a set of 1000 m$^2$ scintillator detectors. The Pierre Auger observatory normalized at 1000 m$^2$ appears 8-9 times more sensitive than scintillators at the same altitude, i.e as sensitive as scintillators at $\sim$3000 m and the 16000 m$^2$ full array is as sensitive as 1000 m$^2$ scintillators at $\sim$5000 m. For this energy cutoff, the SD is about as sensitive as the ARGO YBJ observatory. Above 100 GeV the minimum fluences still decrease but most of the theoretical work (see for instance [11, 12]) on the prompt high energy emission of GRBs, based on the calculation of the synchrotron self Compton component (SSC), predict an energy cutoff well below 1 TeV for reasonable assumptions on the burst parameters. TeV emission predicted in some cases in association with optical flashes [13] could be also observed if the fluences are high enough. As an example the limit fluences on the high energy emission of GRB 970417a (above 50 GeV) given by MILAGRITO [14] are compatible with the SD detection limits and this burst in the field of view of the observatory (below 30° of zenith angle) would have been seen whatever the energy cutoff and the spectral index. Furthermore, in the case of very high energy emission (above 10 TeV) a coincidence between single particles excess and higher level trigger [5] counting rate could be used to remove the traditional degeneracy between the energy cutoff and the spectral index inherent to the single particle technique.

## 5. Discussion and conclusion

Despite its low altitude, the Pierre Auger observatory is a very competitive high energy GRB ground detector for the single particle technique due to its large area and sensitivity to photons. The sensitivity of the Observatory to GRBs is among the best and comparable with ARGO. Therefore interesting constrains on high energy emission of GRBs should be obtained during the next fifteen years of operating time of the Observatory. The potential of water Cherenkov tanks for this kind of detection is clear and one can imagine using this technique, for instance, at the Mount Chacaltaya altitude. At this high altitude the background is roughly 10 times higher than at 1400 m [2] and the signal due to photon shower particles is $\sim$100 times higher (below 10 GeV). The sensitivity of a tank is thus $\sim 100/\sqrt{10} \simeq 32$ times higher at this altitude. An array with a tank size, granularity and water level properly chosen, combining single particle counting, higher level triggers and shower reconstruction could then be a very efficient counterpart to the GLAST satellite observations.